# Short-Term Reliability Evaluation of Generating Systems Using Fixed-Effort Generalized Splitting

Osama Aslam Ansari, *Student Member, IEEE*, S. Mahdi Mazhari, Yuzhong Gong, *Member, IEEE,* and C. Y. Chung, *Fellow, IEEE*

*Abstract*—The short-term reliability evaluation techniques provide a rational approach for risk-informed decision making during power system operation. The existing reliability assessment techniques involve large computational burden and therefore are not directly applicable for short-term reliability evaluation during system operation. To this end, this paper presents a computationally-efficient approach for short-term reliability evaluation of wind-integrated generating systems. The proposed approach makes use of the fixed-effort generalized splitting (FEGS) technique, which is a variant of importance splitting. To realize the implementation of FEGS, a discrete version of component-wise Metropolis-Hastings (MH) algorithm for Markov Chain Monte-Carlo (MCMC) is also presented. Besides, the proposed FEGS approach is extended to take the uncertainties of wind generation and load demand into account. The simulation results indicate that, in comparison to crude Monte-Carlo simulation (CMCS), the proposed approach is able to evaluate short-term reliability indices with a low computational burden. Moreover, further simulation results indicate the impacts of uncertainties of wind generation and load demand on short-term reliability indices.

## I. INTRODUCTION

Power systems are inherently uncertain in nature. The uncertainties in power systems stem from the unexpected or random failures of generating units and transmission lines, sudden changes in the load demand, and unpredictable output from renewable energy sources. Reliability evaluation techniques offer a probabilistic approach to quantifying the impacts of such uncertainties on power systems planning and operation [1], [2]. Although considerable research has been performed in developing and employing long-term reliability techniques for power system planning, the short-term reliability evaluation techniques for power system operation need further research. The development of these methods could allow power system operators to replace the existing heuristic techniques (such as $N-1$ contingency analysis) [3] with fully probabilistic approaches [4], thereby allowing power system operators to take risk-informed decisions amid uncertainties.

This work was supported in part by the Natural Sciences and Engineering Research Council (NSERC) of Canada and the Saskatchewan Power Corporation (SaskPower).

Osama Aslam Ansari*, S. Mahdi Mazhari, Yuzhong Gong, and C. Y. Chung are with the Department of Electrical and Computer Engineering, University of Saskatchewan, Saskatoon, SK, S7N 5A9, Canada (*email: oa.ansari@usask,ca).

The existing short-term reliability evaluation techniques can be classified into three groups: analytical methods, simulation techniques and hybrid approaches. In [5] and [6], a state enumeration approach – an analytical technique, is employed to evaluate short-term reliability of generating systems. Although analytical techniques can precisely calculate the short-term reliability indices, such techniques are computationally-prohibitive for large, realistically-sized power systems [2]. To circumvent this problem, Monte-Carlo simulation (MCS) techniques have been employed [7]–[10]. The extremely poor computational efficiency of crude MCS (CMCS) for rare events (small probability events) is mitigated by different variance reduction techniques (VRTs). For instance, in [7] and [8], importance sampling is adopted to reduce the variance of MCS estimator. Reference [9] proposes a three-stage sequential importance sampling approach as a VRT. In [10], Latin hypercube sampling is combined with Gibbs sampling to consider conditional probability distribution functions (PDFs) while improving the computational performance of MCS. Different from pure analytical and simulation techniques, in [11], a hybrid framework is proposed, where the computational speed is improved by adopting importance sampling.

The importance sampling techniques used previously in [7]–[9], and [11], have certain limitations. The selection of optimal importance sampling density in importance sampling is a key challenge. Sub-optimal importance sampling densities could result in erroneous estimates of reliability indices [12]. Furthermore, for high dimensional cases, importance sampling suffers from degeneracy and may lead to variance explosion [13]. In addition, the variants of importance sampling technique employed in [7], [8] and [11] are generally suitable for PDFs belonging to exponential family, such as beta or normal distributions. The random variables in power systems do not always follow these standard distributions.

Importance splitting is another VRT that does not suffer from the aforementioned drawbacks [14], [15]. Importance splitting sequentially explores the sample space by creating copies of simulation paths as the simulation reaches closer to the rare event [15]. This technique requires no constraints on the definition of PDFs for random variables. In power systems literature, importance splitting has been used for simulation of cascading blackouts [16], [17], where the probability of number of tripped lines are evaluated. However, there is a dearth of literature on the application of importance splitting for short-term reliability evaluation of power systems.

Motivated by aforesaid observation, this paper proposes the application of an efficient importance splitting technique called the fixed-effort generalized splitting (FEGS) for short-term reliability evaluation of generating systems. Compared to other importance splitting techniques, the FEGS can be employed for static, non-Markovian problems [18] and thus is suitable for short-term reliability evaluation of power systems. To implement the FEGS technique, the Metropolis-Hastings (MH) algorithm for Markov Chain Monte-Carlo (MCMC) is modified for discrete random variables of power systems. The framework is also extended to consider the uncertainties of load demand and wind generation. Case studies on the 24-bus IEEE Reliability Test System (RTS) [19] are performed to show the computational superiority of the proposed approach over crude CMCS and the impact of wind and load uncertainties on short-term reliability.

The rest of the paper is organized as follows. In Section II, a brief background of short-term reliability evaluation of generating systems is presented. Section III introduces the importance splitting method. In Section IV, the proposed FEGS approach is presented. Section V performs case studies to explore the efficacy of the FEGS approach. Finally, Section VI provides conclusion of the paper.

## II. BACKGROUND

Consider a power system with $N^G$ number of generating stations. Further, assume that the $g^{th}$ generating station is composed of $N^g$ identical generating units, each with the generating capacity $P^g$. Let $X$ be a discrete random vector indicating the number of available generating units in each generating station.

$$X = \left[n^1, n^2, \ldots, n^{N^G}\right], \quad (1)$$

where $n^g$ is a discrete random variable denoting the number of available generating units at $g$th generating station. Let $S(X)$ be the total available generation capacity of power system during a lead time $\Delta t$. Note that $S(X)$ is also known as the importance function in importance splitting literature [14].

$$S(X) = X(P^G)^T, \quad (2)$$

where $P^G = \left[P^1, P^2, \ldots, P^{N^G}\right]$.

For short-term reliability evaluation, we are interested in finding the probability $P(S(X) \leq L)$, where $L$ is the load demand during the lead time $\Delta t$. This probability, also known as the short-term risk or unit commitment risk $R$, can be evaluated by the following integral:

$$R = P(S(X) \leq L) = \int I\{S(x) \leq L\}f(x)dx, \quad (3)$$

where

$$I\{S(X) \leq L\} = \begin{cases} 0, & S(X) > L \\ 1, & S(X) \leq L \end{cases}, \quad (4)$$

$$f(X) = \prod_{g=1}^{N^G} f_g(n^g). \quad (5)$$

In (5), $f_g(n^g)$ is the binomial distribution with the number of trials parameter equal to $N^g$ and the success probability parameter equal to $1 - \lambda_g \Delta t$; $\lambda_g$ being the outage rate of a generating unit at $g^{th}$ generating station.

In order to evaluate (3), the most common approach is to use a CMCS technique, where samples are drawn from $f(\cdot)$, and (4) is evaluated for each sample. The probability in (3) is then estimated as:

$$\hat{R} = \frac{1}{N}\sum_{i=1}^{N} I\{S(x_i) \leq L\}, \quad (6)$$

where $x_i, \forall i$ are independent and identically distributed samples (IID) from $f(\cdot)$. As $I\{S(x_i) \leq L\}, \forall i$ are Bernoulli random variables, the variance of the estimator in (6) can be analytically evaluated.

$$\text{var}(\hat{R}) = \frac{1}{N}R(1-R). \quad (7)$$

The relative error (RE), which denotes the accuracy of CMCS, can then be evaluated as:

$$\text{RE} = \text{std}(\hat{R})/\mathbb{E}[\hat{R}] = \sqrt{\frac{(1-R)}{RN}}. \quad (8)$$

From (8), it is evident that for a fixed relative error (1-5%), the number of samples required is inversely proportional to the short-term reliability index being estimated. Thus, a large number of samples are required in CMCS for evaluating typical short-term probabilities in the range of $10^{-6}$ to $10^{-3}$. Moreover, (8) also indicates that RE can be decreased by reducing the variance, and in turn, the standard deviation of the estimator. Through importance sampling, it is possible, theoretically, to obtain a different PDF $f^*(\cdot)$ that minimizes this variance to zero. This theoretical PDF $f^*(\cdot)$ is given by

$$f^*(X) = f(X)I\{S(x) \leq L\}/P(S(X) \leq L), \quad (9)$$

and the estimator in (3) is replaced by the following estimator:

$$R = P(S(X) \leq L) = \int I\{S(x) \leq L\}\left(\frac{f(x)}{f^*(x)}\right)f^*(x)dx. \quad (10)$$

Equation (9) represents the main challenge of importance sampling [12], i.e., to find a new PDF $\hat{f}^*(\cdot)$, which is a good approximation of (9).

## III. IMPORTANCE SPLITTING

Importance splitting offers an attractive alternative for importance sampling. Importance splitting is a highly versatile and flexible rare-event simulation technique [14]. The basic idea behind it is to employ sequential sampling to probe regions of sample space which are of interest for rare-event simulation [15]. In short-term reliability evaluation, these regions correspond to failure regions, i.e. regions of system states which lead to load curtailment during the lead time.

The key advantages of importance splitting over importance sampling are two-fold. First, importance splitting does not involve any change of PDFs of the underlying phenomenon, thereby avoiding the drawback of finding importance sampling density. Second, there is no restriction on the definition of $f(\cdot)$ in (3).

Considering the context of short-term reliability evaluation of generating systems, in importance splitting, the sample space is divided by a number of non-identical levels $\{L_0, L_1, \ldots, L_T\}$ with $L_0 > L_1 > \cdots > L_T$. Each level corresponds to a hypothetically higher load demand. The final level $L_T$ is set to the actual load demand. Starting from $L_0$, which is equal to the maximum capacity of the generating system, the stochastic process is simulated until the process returns to the starting point. During this process, the samples that have values of $S(X)$ below $L_1$ are recorded. From each of these samples, multiple new stochastic processes are simulated again until these new processes return to $L_0$. Similarly, further simulations are carried out starting from samples with $S(X)$ values that are below $L_2$. The whole process continues until the region below $L_T$ is not sufficiently explored.

Let $N_t$ be the number of samples that enter into the region $S(X) \leq L_t$ during $t-1$ simulation stage, and let $s_t$ be the splitting factor for the next stage. The conditional probability $P(S(X) \leq L_t | S(X) \leq L_{t-1})$ can be estimated as

$$P(S(X) \leq L_t | S(X) \leq L_{t-1}) = N_t / (N_{t-1} s_t) \quad (11)$$

Through evaluation of these conditional probabilities at different stages or levels, the final probability $P(S(X) \leq L)$ can be evaluated.

Depending on how new processes (or trajectories) are generated, importance splitting has different variants.

### A. Fixed-Splitting (FS)

In FS splitting, at each stage, the number of new trajectories created from the samples that down-crosses $L_t$ is fixed. This number is also known as splitting factor [12]. As the number of new trajectories created is not dependent on the number of samples that down-crosses $L_t$, there is a risk of population explosion [18]. Population explosion implies that the total number of samples at each successive stage grows intractably, which leads to increased computational burden.

### B. Fixed-Effort (FE)

In FE splitting, the total simulation burden at each stage is fixed. This implies that the number of new trajectories (splitting factor) at each stage depends on the number of samples, which down-crosses $L_t$. If there are higher number of such samples, the splitting factor would be lower to keep the total simulated samples fixed. Therefore, the advantage of FE over FS is that the occurrence of population explosion can be avoided.

### C. Fixed Number of Successes (FNS)

In this variant of splitting technique, the simulation at each stage is repeated until a fixed number of samples down-crosses $L_t$. In other words, $N_t$ is fixed for each stage; this approach also avoids the population explosion problem.

## IV. PROPOSED FEGS APPROACH

The importance splitting techniques are typically employed for dynamic, Markovian models [14] and thus cannot be directly adapted to evaluate (3). *Botev* and *Kroese* [18] extended the importance splitting approach to propose GS which could be directly applied to estimate (3). In this paper, it is proposed to evaluate (3) by adapting GS.

### A. Selection of Intermediate Levels

The efficiency of importance splitting is strongly influenced by the choice of intermediate levels $\{L_1, ..., L_T\}$ [14], [15]. These intermediate levels could be selected through an initial run of an importance splitting technique [12]. As proposed in [18], in this paper, the ADAptive Multilevel splitting algorithm (ADAM) is employed to estimate the levels. In essence, the ADAM algorithm employs fixed values of $P(S(X) \leq L_t | S(X) \leq L_{t-1})$ to estimate $\{L_1, ..., L_T\}$ using random populations of samples. Interested readers are referred to [18] for a detailed explanation of the ADAM algorithm.

### B. FEGS Simulation Framework

FEGS is based on the concept of GS. GS extends the traditional splitting method in order for it to be applicable to static, non-Markovian models, such as (3). In essence, the key approach behind GS is to construct a Markov chain at each stage of importance splitting that has a stationary distribution given by:

$$f^t(X) = f(X) I\{S(x) \leq L_t\} / P(S(X) \leq L_t), \quad (12)$$

where, $f^t(X)$ is the distribution for $t^{\text{th}}$ state of importance splitting. In other words, instead of creating multiple new trajectories from each sample that down-crosses $L_t$, multiple Markov chains with the number of samples equal to the splitting factor are generated. The Markov chain can be generated by employing the MCMC techniques having a Markov transition density proportional to $f(X) I\{S(x) \leq L_t\}$. The seeds of these different MCMC are set to the different samples that down-crosses $L_t$. Notice, the similarity between (9) and (12); at the final stage $T$, the Markov chain has a stationary distribution of (7).

Fig. 1 depicts GS for a hypothetical case study. In this figure, three levels are indicated. The initial level $L_0$ represents the total capacity of the generating system, while the final level $L = L_3$ is the load demand. The time axis in Fig. 1 corresponds to the length of Markov chain. In Fig. 1, $N_1$ represents the number of black dots below $L_1$ (2 in this case) and $N_2$ represents the number of grey dots below $L_2$ (4 in this case).

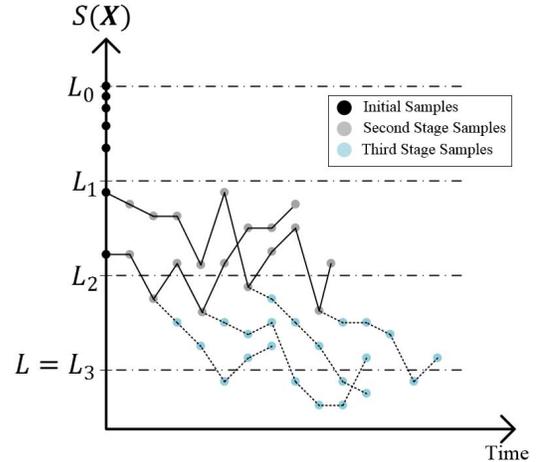

Fig. 1. Pictorial representation of GS.

The GS approach can be employed in both FS and FE approaches. Due to the advantages of FE mentioned previously, in this work, GS is used in conjunction with FE [12], [18]. Algorithm 1 details the FEGS algorithm. The implementation of FE is in step 4 of Algorithm 1, where different splitting factors (Markov chain's lengths) for different entrance samples are generated such that the total expected simulation burden remains $N$ at each stage.

**Algorithm 1** FEGS Algorithm

**Input:** Levels $\{L_1, L_2, ..., L_T\}$, fixed sample size $N$
**Output:** Short-term reliability index in (3)
1:   Set $t = 1$. Sample IID $\{X_1, ..., X_N\}$ from $f(\cdot)$. Set

$$\chi_0 = \{\mathbf{X}_1, \ldots, \mathbf{X}_N\}$$

2: Select $\chi_t = \{\mathbf{X}_1, \ldots, \mathbf{X}_{N_t}\}$ from $\chi_{t-1}$ such that for all samples in $\chi_t$, $S(\mathbf{X}) \leq L_t$.
3: **For** $t = 1$ to $t = T$ **do**
4:     Generate $\{S_t^1, \ldots, S_t^{N_t}\}$, where $S_t^i = \lfloor N/N_t \rfloor + B_i$ and $B_i$ is a Bernoulli random variable with success probability of 0.5, such that $\sum_i B_i = N \bmod N_t$
5:     From each sample $\mathbf{X}_i$ in $\chi_t$, sample $S_{ti}$ IID samples from (12) using MCMC with $\mathbf{X}_i$ as the seed
6:     Collect all samples from step 5 to update $\chi_t$, which results in $N$ samples in $\chi_t$
7:     If $N_t = 0$, set $N_{t+1} = N_{t+2} = \cdots = N_T = 0$
8: **End** the **for** loop
9: Calculate the estimated risk $\hat{R} = N^{-T} \prod_{t=1}^{T} N_t$

### C. MCMC for Discrete PDFs of Generating System

A key step in Algorithm 1 is to generate samples from (12) in step 4. This step could be realized by constructing a Markov chain whose stationary distribution is given by (12). This Markov chain can be generated using MCMC algorithms. In this work, the MH algorithm is adopted for MCMC. The MH-MCMC is inefficient for high dimensional problems [20]. In this paper, a modified version of MH-MCMC is adapted for discrete PDFs of power systems. In this modified version, a new sample is accepted or rejected for each component separately. Algorithm 2 details this component-wise discrete MH-MCMC.

**Algorithm 2** Component-Wise Discrete MH-MCMC

**Input:** Initial sample $\mathbf{X}_0 = \{n_0^1, \ldots, n_0^{N^\mathrm{G}}\}$ following the target distribution $f^t(\cdot)$, and original distribution $f(\cdot)$
**Output:** A population of $N_\mathrm{MC}$ samples $\{\mathbf{X}_1, \ldots, \mathbf{X}_{N_\mathrm{MC}}\}$ following the target distribution $f^t(\cdot)$
1: **For** $i = 1$ to $N_\mathrm{MC}$ **do**
2:    **For** $g = 1$ to $N^\mathrm{G}$ **do**
3:     Draw a candidate sample $\zeta_g$ from a uniform distribution on $\{0, 1, \ldots, N^g\}$
4:     Calculate the acceptance ratio $\alpha = f_g(\zeta_g)/f_g(n_0^g)$
5:     Accept $\zeta_g$ as $n_{i+1}^g$ with probability of $\min\{\alpha, 1\}$ and $n_0^g$ as $n_{i+1}^g$ otherwise
6:    End the **for** loop
7:    Set the proposal sample $\mathbf{X}_p = \{n_{i+1}^1, \ldots, n_{i+1}^{N^\mathrm{G}}\}$
8:    If $\mathbf{X}_p = \mathbf{X}_i$, set $\mathbf{X}_{i+1} = \mathbf{X}_p$ and go to step 10
9:    Evaluate $\mathrm{I}\{S(\mathbf{X}) \leq L_t\}$ using (4) and save the result
10:   **if** $\mathrm{I}\{S(\mathbf{X}) \leq L_t\} = 1$ set $\mathbf{X}_{i+1} = \mathbf{X}_p$, **else** set $\mathbf{X}_{i+1} = \mathbf{X}_0$
11:   Set $\mathbf{X}_0 = \mathbf{X}_{i+1}$
12: **End** the **for** loop

### D. Inclusion of Uncertainties of Load and Wind Generation

The FEGS framework presented in Section IV.B is developed by considering the random outages of conventional generating stations. The uncertainties of load demand and wind generation during the lead time also impacts the short-term reliability of power systems. The previously proposed framework can be extended to include the load uncertainty and wind generation uncertainty. In particular, the definition of $f(\cdot)$ is modified to include the PDF for load demand $f_\mathrm{L}(\cdot)$ and wind generation $f_\mathrm{W}(\cdot)$ during the lead time

$$f(\mathbf{X}) = \prod_{g=1}^{N^\mathrm{G}} f_g(n^g) f_\mathrm{L}(X_\mathrm{L}) f_\mathrm{W}(X_\mathrm{W}), \quad (13)$$

where, $X_\mathrm{L}$ and $X_\mathrm{W}$ are the continuous random variable for load demand and wind generation, respectively, during the lead time, and $\mathbf{X} = \{n^1, \ldots, n^{N^\mathrm{G}}, X_\mathrm{L}, X_\mathrm{W}\}$. The definition of $S(\mathbf{X})$ is also modified as:

$$S(\mathbf{X}) = \mathbf{X}(\mathbf{P}^\mathrm{G})^\mathrm{T} - X_\mathrm{L} + X_W. \quad (14)$$

For simplicity, in this paper, $f_\mathrm{L}(\cdot)$ and $f_W(\cdot)$ are modeled using Gaussian distributions centered at the load demand forecast and wind generation forecast, respectively. As mentioned in the introduction, unlike certain importance sampling techniques, there is no restriction on the choice of $f_\mathrm{L}(\cdot)$ and $f_\mathrm{W}(\cdot)$.

Fig. 2 pictorially represents the FEGS approach when only the load uncertainty is considered. Notice that the final level is always set to zero in this case.

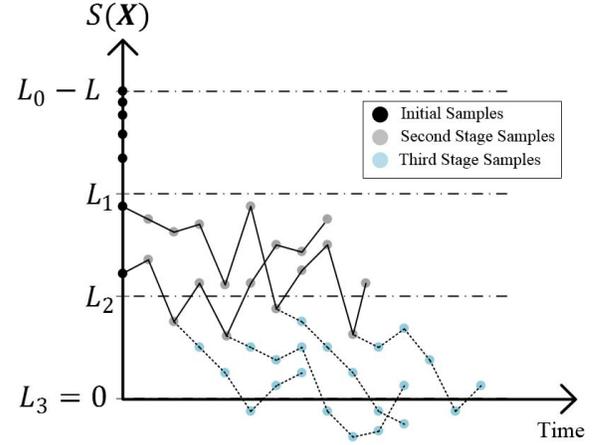

Fig. 2. Pictorial representation of FEGS with load uncertainty.

## V. RESULTS

The efficacy of the proposed FEGS approach for short-term reliability evaluation of wind-integrated power systems is numerically demonstrated. The 24-bus IEEE RTS is employed, which has a total generation capacity of 3,405 MW. The fixed sample size $N$ in Algorithm 1 is set between 10,000 to 150,000. Higher values of $N$ are used when the estimated short-term risk indices are expected to have lower values. All simulations are performed on a personal computer with a 3.40 GHz Intel® Core i7-4770 CPU and a 16 GB RAM. MATLAB is used to implement the proposed framework.

### A. Demonstrative Case

In this section, the computational efficiency of the proposed FEGS approach over CMCS is demonstrated. The stopping criteria for CMCS is based on RE and is set to 10%. Table I compares the computational performance of the two approaches. $N_\mathrm{MCS}$ represents the number of times $S(\mathbf{X})$ is evaluated. Higher $N_\mathrm{MCS}$ corresponds to higher computational time. As it is evident from Table I, the proposed FEGS approach achieves superior computational performance compared to the CMCS. Moreover, the computational superiority of FEGS over CMCS increases significantly with the decrease in the short-term risk index. In Table I, the slight deviation between the risk indices estimated by the two approaches is due to the fact to 10% RE in CMCS estimations.

TABLE I
COMPUTATIONAL PERFORMANCE OF FEGS-MCS VS. CMCS

| Load ($L$) (MW) | CMCS | | FEGS-MCS | |
|---|---|---|---|---|
| | Risk ($10^{-5}$) | $N_{MCS}$ | Risk ($10^{-5}$) | $N_{MCS}$ |
| 3100 | 540.31 | 21,500 | 561.24 | 23,000 |
| 3000 | 2.4964 | 431,400 | 2.3724 | 33,020 |
| 2900 | 7.1920 | 1,220,700 | 7.4240 | 35,568 |
| 2850 | 3.4896 | 3,250,800 | 3.2005 | 34,238 |
| 2700 | 1.7853 | 5,000,000* | 1.7018 | 39,903 |

* Maximum number of evaluations for MCS was reached and RE was 13%.

### B. Impact of Load Uncertainty

In this section, the uncertainty of load during the lead time is also considered. The load demand forecast value is set to 2,850 MW. Table II shows the short-term reliability indices for different standard deviations of the Gaussian distribution for load demand. As expected, the results indicate that the increase in uncertainty corresponds to lower short-term reliability. A comparison of Table I with Table II indicates an increase in computational burden when load uncertainty is included.

TABLE II
SHORT-TERM RELIABILITY INDICES CONSIDERING LOAD UNCERTAINTIES

| Standard Deviation (% of load) | FEGS-MCS | |
|---|---|---|
| | Risk ($10^{-5}$) | $N_{MCS}$ |
| 0.1 | 3.5344 | 97,736 |
| 0.5 | 3.9089 | 95,374 |
| 1 | 5.5137 | 92,131 |
| 2 | 6.6393 | 81,337 |
| 3 | 17.7350 | 88,510 |
| 5 | 82.407 | 77,961 |

### C. Impact of Wind Generation Uncertainty

In this section, the impact of both load and wind generation uncertainty is studied. The standard deviation for load uncertainty is fixed to 0.1% of the forecast load demand. The standard deviation for wind generation PDF is set to 10% of forecast value. Two cases are considered.

**Case A**: A wind farm is committed and a 155 MW generating station at bus 15 is de-committed.

**Case B**: Only conventional generators are committed and wind generation is not used.

Table III reports the results for this case study. The results indicate that, when wind generation with a forecast value of 155 MW is committed, the short-term reliability of the generation system drops. Although the maximum generation capacities of the system for both cases are identical, the uncertainty associated with wind generation increases the risk of load curtailment, thereby reducing the short-term reliability. However, at higher wind generation forecast values, the short-term reliability improves due to additional available generation.

TABLE III
SHORT-TERM RELIABILITY INDICES CONSIDERING WIND GENERATION

| Case | Wind Generation Forecast | | |
|---|---|---|---|
| | 155 MW | 200 MW | 300 MW |
| Case A | $4.8159 \times 10^{-5}$ | $2.0071 \times 10^{-5}$ | $1.0817 \times 10^{-5}$ |
| Case B | | $3.5344 \times 10^{-5}$ | |

## VI. CONCLUSION

In this paper, a new approach for short-term reliability evaluation of generating systems is proposed. The proposed approach employs FEGS, a computationally-efficient MCS technique. A discrete version of component-wise MH-MCMC is presented to implement the FEGS approach. The results have shown the computational superiority of the proposed approach over CMCS. Further simulation results have indicated the impact of uncertainties of load and wind on short-term reliability indices. The method developed in this paper could be utilized by power system operators for risk-informed decision-making during system operation, such as during unit commitment and economic dispatch.